\documentclass[preprint,trackchanges]{aastex6}
\usepackage{xspace}
\definecolor{DarkGreen}{rgb}{0.0,0.4,0.0}  

\newcommand{\kms}{km~s$^{-1}$\xspace}
\newcommand{\alfven}{Alfv\'{e}n\xspace}
\newcommand{\rsun}{$R_\odot$\xspace}
\begin{document}


\title{Unraveling the Links among Sympathetic Eruptions}


\author{Dong Wang\altaffilmark{1,2}, Rui Liu\altaffilmark{1}, Yuming Wang\altaffilmark{1}, Tingyu Gou\altaffilmark{1}, Quanhao Zhang\altaffilmark{1}, Zhenjun Zhou\altaffilmark{3}, Min Zhang\altaffilmark{2}}


\altaffiltext{1}{CAS Key Laboratory of Geospace Environment, Department of Geophysics and Planetary Sciences, University of Science and Technology of China, Hefei, Anhui 230026, China; rliu@ustc.edu.cn}
\altaffiltext{2}{Department of Mathematics and Physics, Anhui Jianzhu University, Hefei, Anhui 230601, China}
\altaffiltext{3}{School of Atmospheric Sciences, Sun Yat-sen University, Zhuhai, Guangdong 519000, China}

\begin{abstract}
Solar eruptions occurring at different places within a relatively short time interval are considered to be sympathetic. However, it is difficult to determine whether there exists a cause and effect between them. Here we study a failed and a successful filament eruption following an X1.8-class flare on 2014 December 20, in which slipping-like magnetic reconnections serve as a key causal link among the eruptions. Reconnection signatures and effects are: at both sides of the filament experiencing the failed eruption, serpentine ribbons extend along chromospheric network to move away from the filament, while a hot loop apparently grows above it; at the filament undergoing the successful eruption, overlying cold loops contract, while coronal dimming appears at both sides even before the filament eruption. These effects are understood by reconnections continually transforming magnetic fluxes overlying one filament to the other, which adjusts how the magnetic field decays with increasing height above the filaments in opposite trends, therefore either strengthening or weakening the magnetic confinement of each filament.

\end{abstract}




\section{Introduction} \label{sec:intro}
The most energetic phenomena in the solar atmosphere are flares, filament eruptions, and coronal mass ejections (CMEs). CME-caused disturbances and restructuring are often global \citep[e.g.,][]{Hudson1996}. Naturally the impact from a CME is expected to trigger the eruption of a nearby structure already on the verge of unstableness. But in practice it is hard to establish causal links between successive eruptions. This has been a key issue in research on sympathetic eruptions, eruptions that take place nearly synchronously in separated regions on the sun. Here the synchronicity is not meant in its literal sense but allows a short temporal separation of no more than a few hours \citep{Moon2003}. 

The separated regions involved in sympathetic eruptions could be connected by large-scale coronal loops \citep[e.g.,][]{Wang2001,Jiang2008} or by large-scale magnetic skeletons such as separators, separatrices, and quasi-separatrix layers \citep{Schrijver+Title2011,Titov2012,Schrijver2013}. The long-distance coupling may result from the impact of CMEs, waves, or propagating perturbations \citep[e.g.,][]{Wang2001,Jiang2008,Torok2011,Lynch+Edmondson2013,Jin2016}. A crucial physical link between sympathetic eruptions is often proposed to be magnetic reconnection of large-scale magnetic field as induced indirectly or directly by distant or nearby eruptions: the reconnection may effectively modify the overlying field that provides the confining force for a pre-eruptive structure underneath, as suggested in numerous observational investigations \citep{LiuC2009, Zuccarello2009, Jiang2011, Shen2012, Yang2012, Joshi2016, Wang2016} and corroborated by a few numerical simulations \citep{Ding2006,Torok2011,Lynch+Edmondson2013}. The consequence of the reconnection is occasionally observed as coronal dimming \citep{Jiang2011} when the overlying field is partially removed by reconnection. However, smoking-gun evidence for reconnections of large-scale magnetic field has seldom been reported in the literature, supposedly because such reconnections involve weaker magnetic field and are hence much less energetic than those in major flares and CMEs, therefore leaving only elusive signatures in the solar atmosphere. 

In this paper, we present unambiguous evidence for magnetic reconnection of large-scale magnetic field involved in a set of sympathetic eruptions, including an X-class flare and its accompanying CME, a failed and a successful filament eruption.  We conclude that the reconnection plays an important role in regulating the behavior of the filament eruptions. In the sections that follow, we analyze the observations in Section 2, and interpret the observations in Section 3.

\section{Observation \& Analysis}\label{sec:calculate}

\subsection{Overview}
In this study we mainly used EUV images taken by the Atmospheric Imaging Assembly \citep[AIA;][]{lemen2011atmospheric} on-board the Solar Dynamics Observatory \citep[SDO;][]{Pesnell2012}. AIA provides high resolution ($1.5''$) and high cadence (12 s) full-disk images up to 1.28 R$_\sun$ at multiple passbands. AIA's EUV passbands cover a wide temperature range. Among them, 131, 94, and 335~{\AA} are usually sensitive to hot plasma in active regions; 171, 193, and 211~{\AA} respond well to ``quiet'' coronal loops; 304~{\AA} mainly covers the chromosphere and transition region. However, caution has to be taken because each passband contains multiple emission lines \citep{Boerner2012}. 
  
The set of sympathetic eruptions under investigation include an X1.8-class flare from the NOAA active region (AR) 12242, a failed and a successful filament eruption originating from a quiet region located to the south of AR 12242 (Figure~\ref{fig:overview}). The X1.8-class flare starts at 00:11 UT and peaks at 00:28 UT on 2014 December 20 (Figure~\ref{fig:timeline}a). In AIA 131~{\AA}, an expanding bubble-like structure can be seen to erupt southwestward from AR 12242 (Figure~\ref{fig:overview}c and accompanying animation). The eruptive structure make different effects on three nearby filaments, one is located at the south boundary of the active region (F1 hereafter) and the other two are quiescent filaments (F2 and F3 hereafter) located to the further south (Figure~\ref{fig:overview}(a \& b)). F1 is close to the flare site, but remains intact during the flare. F2 is apparently disturbed by the eruptive bubble, but it fails to erupt. F3 initially survives the X-class flare, but becomes unstable later and erupts during the decay phase of the flare. 

To establish the timeline, we constructed a time-distance map (Figure~\ref{fig:timeline}b) to show the ejection associated with the X-class flare seen through a virtual slit S1 (Figure~\ref{fig:overview}c), and a similar map (Figure~\ref{fig:timeline}c) to show the successful filament eruption seen through a virtual slit S2 (Figure~\ref{fig:overview}d). With the soft X-ray 1--8~{\AA} flux given by Geostationary Operational Environmental Satellite (GOES; Figure~\ref{fig:timeline}a), one can see that the earlier eruption is synchronized with the X-class flare, but the later one is delayed by half an hour. Employing the Large Angle Spectroscopic Coronagraph (LASCO) on-board the Solar and Heliospheric Observatory (SOHO), we identified both the earlier CME associated with the X-class flare (Figure~\ref{fig:overview}e) and the later one resulting from the successful filament eruption (Figure~\ref{fig:overview}f, see also Figure~\ref{fig:timeline}a). As the later CME closely trails the earlier one, they are registered as one single event in the SOHO LASCO CME catalog. The two CMEs occur well within the temporo-spatial range that is determined by the \alfven speed, typically on the order of 1000~\kms in the corona, since the time difference between the onset of the X-class flare and F3's eruption is about 40 min, and the distance from the flare to F3 is about 400 Mm, as measured along the spherical surface (Figure~\ref{fig:overview}a). Thus a propagation speed over 170~\kms is required for the agent coupling the two events. Since this speed is much smaller than the typical \alfven speed or the earlier CME's speed ($\sim\,$800~\kms; Figure~\ref{fig:timeline}a), it is unlikely that the later CME is directly caused by the earlier one.

\subsection{Magnetic Configuration} \label{subsec:mag}
 	
To understand the large-scale magnetic configuration, we utilized the \texttt{pfss} package in SolarSoftWare, which reconstructs the coronal field with a potential-field source-surface model \citep[PFSS;][]{Schatten1969}. To approximate the evolving field on the full sphere, the lower boundary of the PFSS model, i.e., the synoptic map of the photospheric magnetic field, is updated every 6 hours by assimilating magnetograms into a flux-dispersal model \citep{schrijver2003photospheric}. 

With a pre-flare PFSS model of the coronal magnetic field (Figure~\ref{fig:pfss}), we calculated the decay index, $n=-d\ln B/d\ln h$, right above the three filaments by manually picking points along each filament visible in 304~{\AA} (Figure~\ref{fig:overview}(a \& b)). The decay index describes how fast the magnetic field decreases with increasing height. \citet{kliem2006torus} found that a toroidal flux ring is unstable to lateral expansion if the decay index of the external poloidal field $B_{\mathrm{ex}}$ exceeds 3/2 (hence the term \emph{torus instability}). Usually one cannot decouple $B_{\mathrm{ex}}$ from the flux-rope field, but can approximate $B_{\mathrm{ex}}$ with a potential field \citep[e.g.,][]{torok2007numerical,Demoulin+Aulanier2010,wang2017critical}. This is because the external field component orthogonal to the flux-rope axis effects the downward $\mathbf{J}\times\mathbf{B}$ force, and potential field is almost orthogonal to the polarity inversion line (PIL), with which the flux rope in equilibrium is typically aligned. Hence in our calculation $n=-d\ln B_h/d\ln h$, where $B_h$ is the horizontal component perpendicular to the radial component $B_r$ in spherical coordinates.

Figure~\ref{fig:decay_index} shows the variation of decay index as a function of height averaged for the selected points of each filament, and the critical height  $h_\mathrm{crit}$ \citep{wang2017critical} corresponding to the theoretical threshold ($n_\mathrm{crit}=1.5$) of the torus instability. One can see that all three filaments are quite stable to the torus instability, with $h_\mathrm{crit}$ exceeding 200 Mm ($\sim\,$0.3~\rsun) for the two quiescent filaments F2 and F3 and exceeding 100 Mm for the AR filament F1. As a comparison, using triangulation methods, \citet{Xu2010} and \citet{Liu2012dd} found a few AR filaments lying below 35 Mm (0.05~\rsun) before eruptions; \cite{Liu2013} found a few quiescent filaments lying below 70 Mm. These numbers agree with a statistics showing that eruptive prominences typically become unstable in a height range 0.06--0.14~\rsun \citep{LiuK2012}.

In particular F1 has a saddle-like $n(h)$ profile, which may provide additional confinement at altitudes corresponding to small decay indexes at the bottom of the saddle  \citep{wang2017critical}. We also calculated F1's decay index in a potential field based only on $B_z$ of the active region, provided by \texttt{hmi.sharp\_cea} data series from Helioseismic and Magnetic Imager \citep[HMI;][]{schou2011design,scherrer2012helioseismic} on-board SDO, but found that $n(h)$ increases monotonously with height (red curve in the left panel of Figure~\ref{fig:decay_index}); this approach gives $h_\mathrm{crit}=61$ Mm, still quite large for an AR filament. The difference between the two $n(h)$ profiles for F1 demonstrates that the high-altitude magnetic field is strongly modulated by the quiet-Sun field outside the active region. The preflare background field may account for F1 and F2's behavior in response to the X-class flare, but cannot explain F3's successful eruption which is significantly delayed relative to the flare. Below we investigate the physical processes linking the flare and the filament eruptions. 

\subsection{Slipping-Like Magnetic Reconnection} 
  
An outstanding feature that precedes F3's eruption is a serpentine brightening on the surface propagating from F2 towards F3 in AIA 304~{\AA} (Figure~\ref{fig:AIA304_1600_slipping} and accompanying animation), which apparently triggers the successful filament eruption when it approaches F3. Comparing AIA 304 and 1600{\AA} images, one can see that the serpentine ribbon slips along the chromospheric network, where the magnetic field (plage field hereafter) is generally stronger than the intranetwork field. This indicates that magnetic field plays a key role here, unlike the surface brightening caused by filament material draining under gravity \citep[e.g.,][]{Gilbert2013}. A similar serpentine ribbon slips northward from F2 toward F1, when F2 becomes fractured in the middle with material moving towards both ends (Figure~\ref{fig:AIA304_slipping} and accompanying animation). F2 disappears several hours later due presumably to the material drainage to the surface. Meanwhile, a weak brightening point on the southern side of F3 appears to extend northward into a ribbon-like feature (bottom panels of Figure~\ref{fig:AIA304_slipping}).

In tandem to the extension of the serpentine ribbon toward F3, a loop overlying F2 appears at about 00:37 UT in 131~{\AA}, apparently rising and expanding, as demonstrated by the snapshots in Figure~\ref{fig:loop}(a--c) and accompanying animation, as well as the stack plot (Figure~\ref{fig:timeline}d) generated by a virtual slit S3 across the loop (Figure~\ref{fig:loop}b). The loop's southern leg is less visible than its northern counterpart, and most of the time both footpoints cannot be clearly discerned, only at 00:40:20 UT can its southern footpoint be identified as a brightening feature to the north of F3 (marked by a red arrow in Figure~\ref{fig:loop}a). This loop is diffuse and only visible in 131~{\AA}, suggesting that it is heated to as hot as $\sim\,$10 MK. It continues to be visible and rising until F3's eruption.  

In contrast to the hot loop, a bundle of large-scale cold loops in 171~{\AA} overlying both F2 and F3 are observed to contract toward F3 when F3 begins to erupt (Figure~\ref{fig:loop}(d--f) and accompanying animation). In view of the PFSS model (Figure~\ref{fig:pfss}), this loop bundle corresponds to the large-scale arcade that overarches both the active region and the two quiescent filaments. It is anchored in the negative-polarity plage to the southeast of AR 12242 and the positive-polarity plage to the west of AR 12242 (see also Figure~\ref{fig:overview}a). The stack plot generated by the virtual slit S4 across these loops gives a contraction speed of $30\sim40$ km~s$^{-1}$ (Figure~\ref{fig:timeline}e). 

The cold loop contraction and hot loop expansion during about 00:40--01:00 UT are also associated with coronal dimming which extends southward from the active region toward F3 (Figure~\ref{fig:AIA_base_diff}). Visible in base-difference images of all three AIA passbands sensitive to the quiet corona, i.e., 171, 193, and 211~{\AA}, the dimming indicates mass evacuation. Starting at about the flare peak time (Figure~\ref{fig:timeline}), the dimming may initially be associated with the eruption from the active region, but later it extends over F3 to its southern side (Figure~\ref{fig:AIA_base_diff} and accompanying animation). Before F3's eruption, the rapid decrease in brightness (Figure~\ref{fig:timeline}f) at two representative dimming points on both sides of F3 (marked by a red and green cross in Figure~\ref{fig:AIA_base_diff}) is temporally associated with the ongoing expansion of hot loops (Figure~\ref{fig:timeline}d) and contraction of cold loops (Figure~\ref{fig:timeline}e). 

\section{Discussion \& Conclusion}                                     
\begin{figure}[ht!]
 \plotone{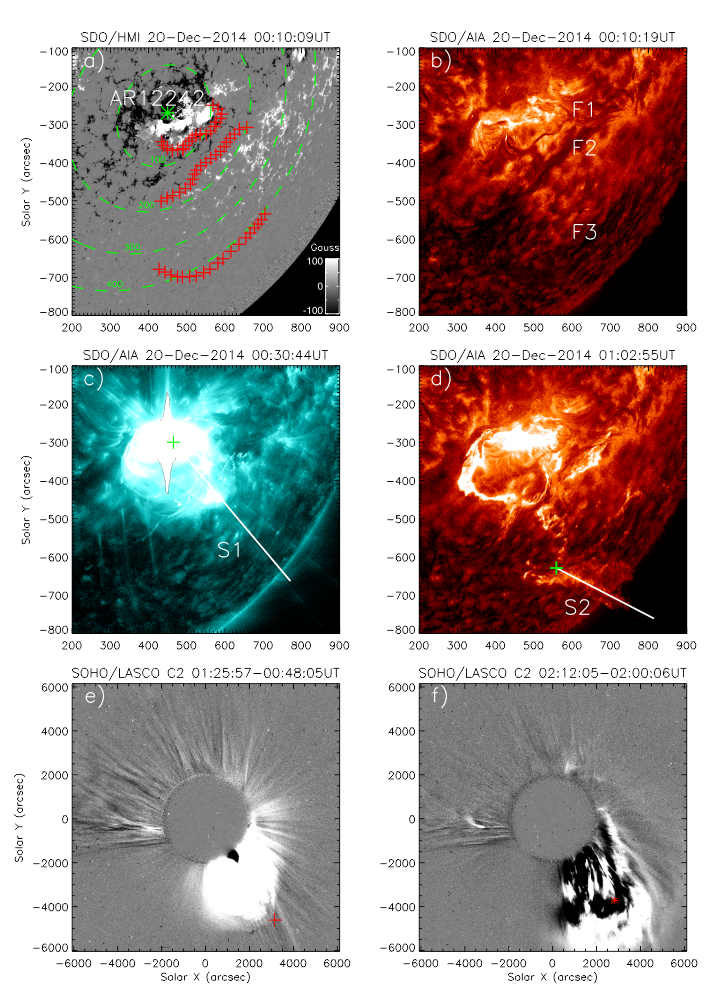}
 \caption{Overview of the sympathetic eruptions. a) shows the southwest quadrant of a line-of-sight HMI magnetogram before the X1.8-class flare. Red plus symbols mark the locations of the three filaments, F1, F2, and F3, in the corresponding AIA 304~{\AA} image in b). Green dashed circles indicate the distance measured along the surface of the sphere from the center of the active region (marked by a green asterisk) in units of Mm. The flare is shown in AIA 131~{\AA} in c), and F3's eruption in AIA 304~{\AA} in d). Their corresponding CMEs are observed by LASCO C2 in e) and f), respectively, with the former CME front marked by a red plus, and the later by a red asterisk. Two virtual slits, S1 and S2, are indicated in (c) and (d), respectively,  and their starting points are marked by green plus symbols. An animation of AIA 131 and 304~{\AA} images is available online.  \label{fig:overview}}
\end{figure}

\begin{figure}[ht!]
	\plotone{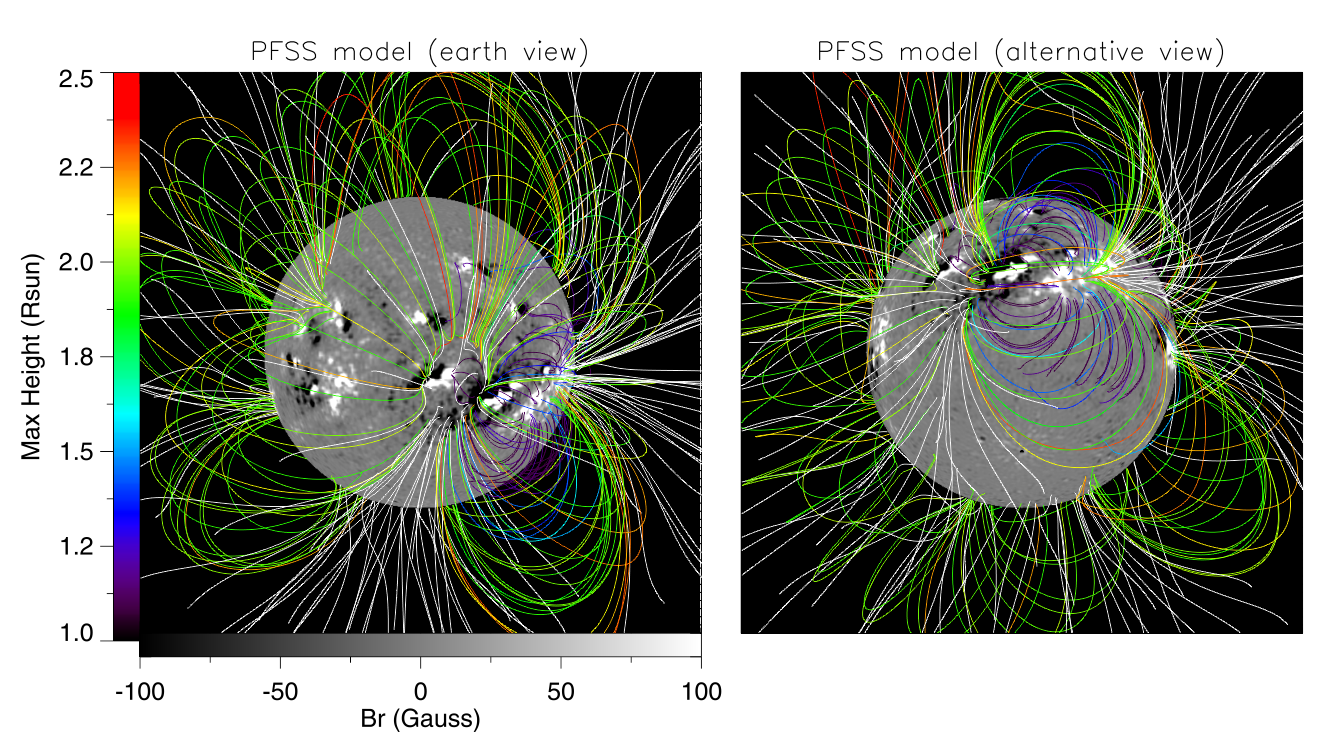}
	\caption{Coronal magnetic field reconstructed with the PFSS model. The plots combine the large-scale field lines around the globe and small-scale field lines in the southwest quadrant where AR 12242 is located. Open field lines are shown in white. Closed field lines are colored by each own maximum height, and viewed from the Earth (left) and from a southern perspective (right). \label{fig:pfss}}
\end{figure}

\begin{figure}[ht!]
	\plotone{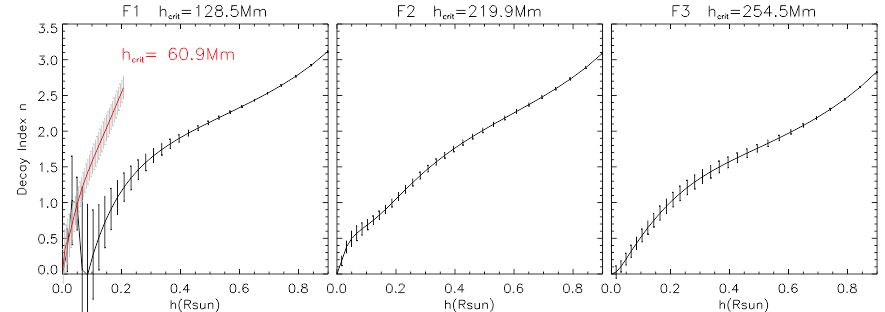}
	\caption{Decay index as a function of height above the three filaments of interest. The height is given in units of $R_{sun}$ above the surface. For F1, the decay index is calculated with both a PFSS model (black) and a local potential extrapolation (red). The error bars result from standard deviation of the manually picked points shown in Figure~\ref{fig:overview}.  \label{fig:decay_index}}
\end{figure}

\begin{figure}[ht!]
 \plotone{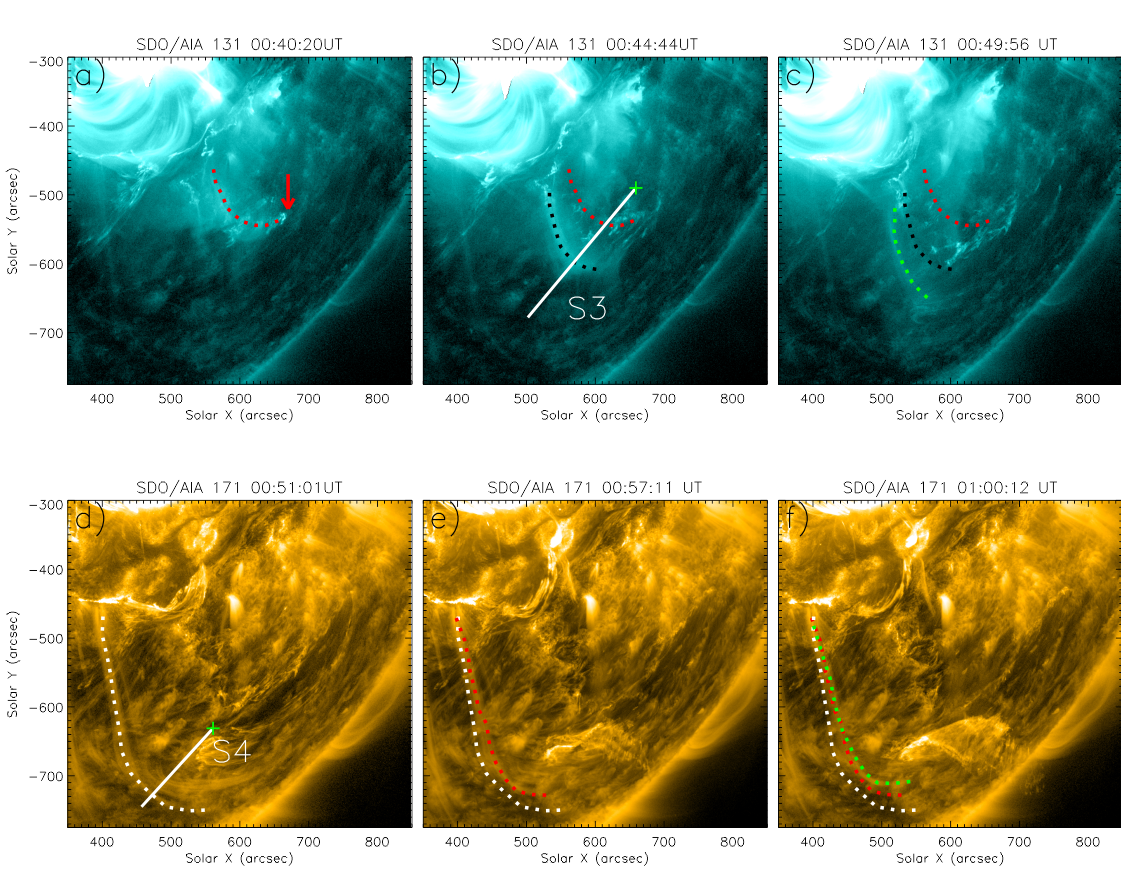}
 \caption{Loop dynamics associated with the slipping-like reconnection. (a--c) Loop rising and expansion observed in AIA 131~{\AA}. The red, black and green dotted curves delineate the loop shape at 00:40, 00:44, and 00:49 UT, respectively. The red arrow in (a) marks the footpoint brightening of the loop. (d--f) Loop contraction observed in AIA 171~{\AA}. The white, red and green dotted curves delineate the loop shape at 00:51, 00:57, and 01:00 UT, respectively. Two virtual slits, S3 and S4, are indicated in (b) and (d), respectively, and their starting points are marked by green plus symbols. An animation of A1A 131 and 171~{\AA} images is available online. \label{fig:loop}}
\end{figure}

\begin{figure}[ht!]
 \plotone{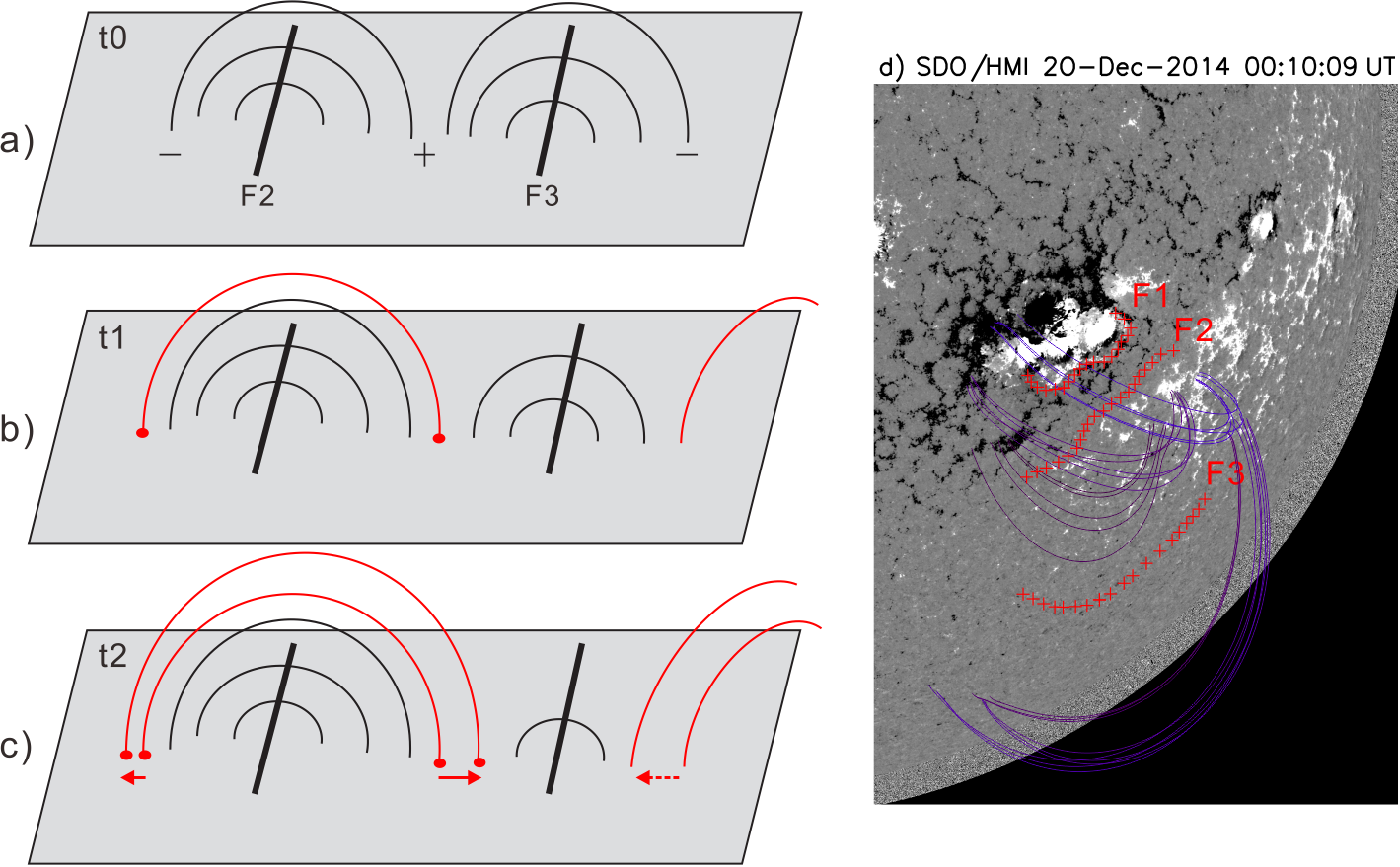}
 \caption{Schematic diagram of the slipping-like magnetic reconnection. a--c) The black thick lines indicate the filaments F2 and F3. The plus and minus signs denote positive and negative polarity, respectively. Magnetic field lines are denoted by thin lines, with black and red colors indicating those before and after reconnection, respectively. The red arrows mark the slipping directions observed on the surface as field lines are `lightened' successively by magnetic reconnection. The red dots mark the footpoints of newly reconnected, closed field lines, indicating the brightening in the chromosphere and transition region. d) Representative PFSS field lines traced from the positive-polarity plage field between F2 and F3. The field lines are color-coded in the same way as in Figure~\ref{fig:pfss}. \label{fig:cartoon}}
\end{figure}

To account for the observational features presented above, a schematic diagram is given in Figure~\ref{fig:cartoon}, based on the well-known fact that filaments are aligned along polarity inversion lines. For simplicity we have omitted AR 12242 and F1 but focus on F2 and F3, conjecturing that F1 lies low under strong magnetic confinement (see \S\ref{subsec:mag}) so that it is not affected by the CME-associated disturbances from the active region \citep[e.g.,][]{Liu2013}. We have also neglected four complications. First, the positive-polarity field between F2 and F3 belongs to a large tract of plage field that extends to far west (Figure~\ref{fig:overview}a). This plage field, together with the negative-polarity plage field to the southeast of AR 12242, dominates the large-scale magnetic connections in the southwest quadrant of the Sun, as demonstrated by the PFSS model (Figure~\ref{fig:pfss}) and the large-scale loops in AIA 171~{\AA} (Figure~\ref{fig:loop}(d--f)). Second, the field to the south of F3 is very weak and may possess severely mixed polarities, because the south polar field reversed signs i.e., became negative, only nine months before \citep{Sun2015}. Third, F2 and F3 are not parallel and F3 is curved. Fourth, white-light coronagraph images show a large-scale structure fanning out over the southwest quadrant with numerous plasma sheet extensions and persisting for many days (not shown), indicating ample presence of open field. But in the PFSS model, only a few open field lines (white) originate from the far west (Figure~\ref{fig:pfss}). Thus, the actual magnetic configuration must be much more complicated than that depicted in the cartoon.  

\begin{figure}[ht!]
	\epsscale{0.8}
	\plotone{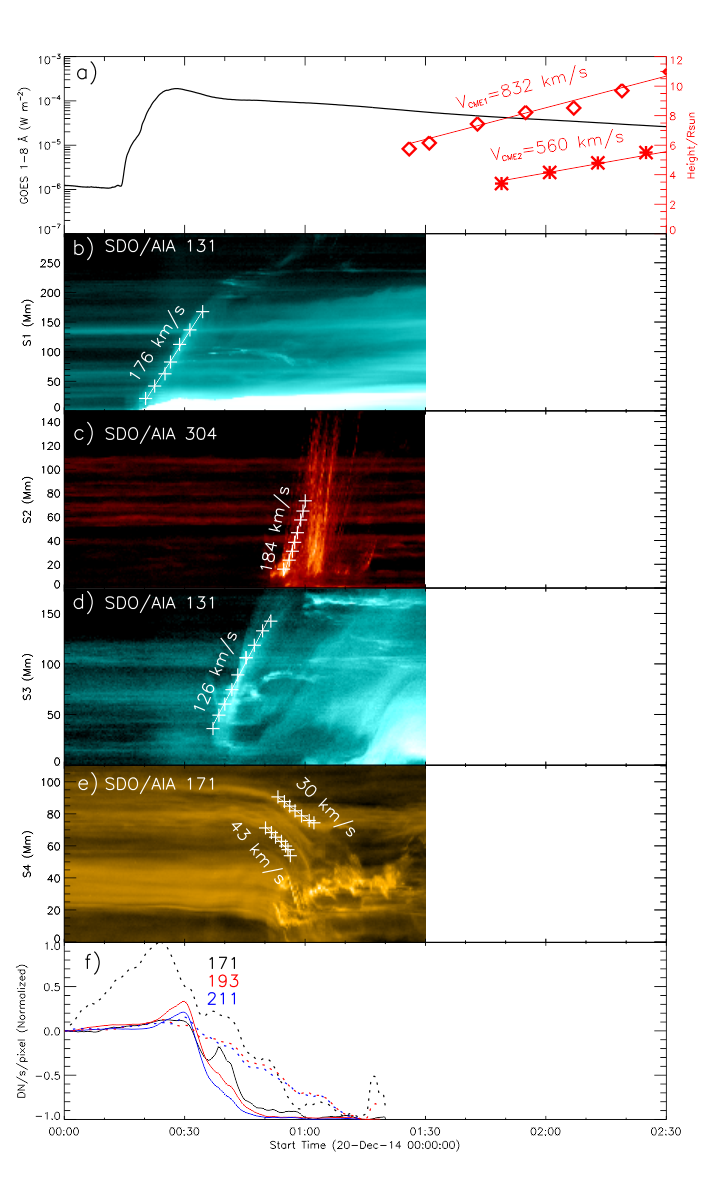}
	\caption{Timeline of the sympathetic eruptions. Panel a) shows \emph{GOES} 1--8~{\AA} light curve, scaled by the left $y$-axis; and heights of the two successive CMEs as a function of time, measured with LASCO's C2 and C3 cameras and scaled by the right $y$-axis. Panels (b--e) show stack plots constructed with the virtual slits S1--S4 as indicated in Figures~\ref{fig:overview}c, \ref{fig:overview}d, \ref{fig:loop}b, and \ref{fig:loop}d, respectively. Linear fittings of the features marked by white crosses in the stack plots give the projected speeds of the eruptive structure associated with the X1.8-class flare (b), the filament eruption (c), the apparent rising and expansion of the hot loop (d), and the contraction of the cold loops (e). Panel (f) shows the variation of normalized mean brightness in two $20''\times20''$ representative dimming regions marked in the bottom-middle panel of Figure~\ref{fig:AIA_base_diff}. The solid (dotted) lines correspond to the region on the northern (southern) side of F3. \label{fig:timeline}}
\end{figure}

\begin{figure}[ht!]
 \plotone{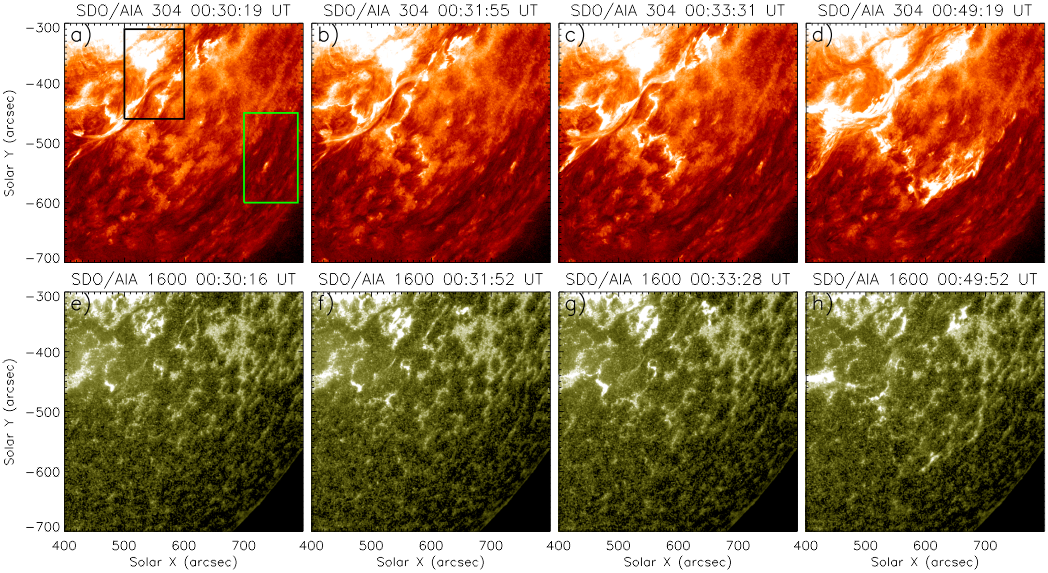}
 \caption{Slipping extension of serpentine ribbons at both sides of F2. A black box marks the brightening to the north of F2, a green box marks a weak brightening to the south of F3. Figure~\ref{fig:AIA304_slipping} zooms in on both regions. An animation of AIA 1600 and 304~{\AA} images is available online, with corresponding base-difference images.  \label{fig:AIA304_1600_slipping}}
 \end{figure}

\begin{figure}[ht!]
 \plotone{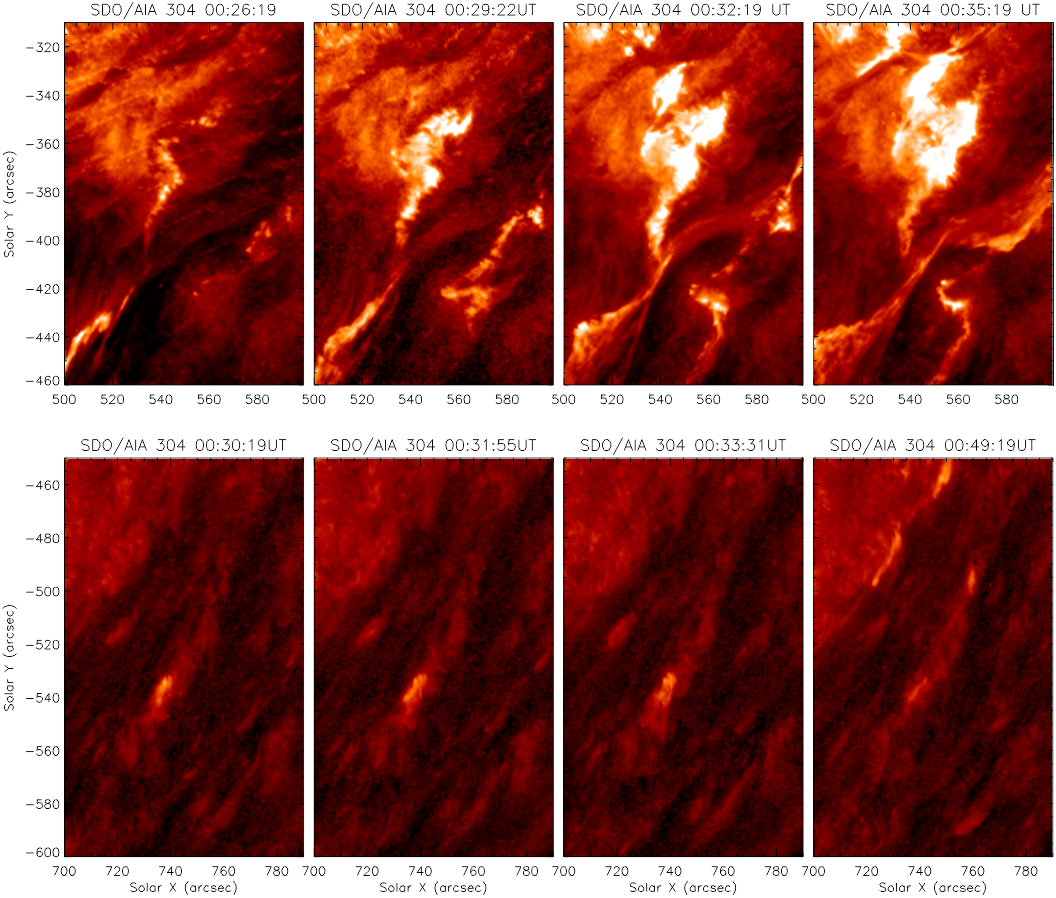}
 \caption{Slipping brightening at far sides of F2 and F3. Top panels show the slipping extension of a serpentine ribbon to the northward of F2. Bottom panels show that a weak brightening to the south of F3 extends northward. The top (bottom) panels' field of view is indicated by the black (green) rectangle in Figure~\ref{fig:AIA304_1600_slipping}a. An animation of AIA 304~{\AA} images with a larger FOV is available online to show F2's failed eruption.  \label{fig:AIA304_slipping}}
\end{figure}

With the above simplifications, the initial magnetic configuration is locally tripolar (Figure~\ref{fig:cartoon}a), reminiscent of a pseudo streamer that overlies two loop arcades side by side, with each occasionally harboring a filament \citep{Wang2007,Torok2011}. Magnetic reconnection in such a topology often involve the quasi-separatrix layer separating the two arcades \citep[e.g.,][]{Liu2014,Gou2016}. Similarly in our case, one can see that PFSS field lines traced from the positive-polarity plage field between F2 and F3 are mapped to distant places, overlying F2 and F3 separately (Figure~\ref{fig:cartoon}d), which indicates the presence of quasi-separatrix layers \citep{Demoulin2006}. Here magnetic reconnection is speculated to be triggered by the CME associated with the X-class flare and to occur between field lines overlying F3 and some higher field lines. The reconnection produces new field lines (red) overlying F2 and those anchored on the right (southern) side of F3, the latter of which are either open, if the higher field lines are open \citep[e.g.,][]{Torok2011}; or connected to remote places, most likely to the positive-polarity plage in the far west, if the higher field lines are closed. Similar reconnections take place successively (Figure~\ref{fig:cartoon}(b \& c)), manifesting themselves as the slipping extension of serpentine ribbons along the chromospheric network away from both sides of F2 (Figures~\ref{fig:AIA304_1600_slipping} and \ref{fig:AIA304_slipping}) and the apparent rising and expansion of hot loops overlying F2 in 131~{\AA} (Figures~\ref{fig:timeline}d and \ref{fig:loop}(a--c)). A similar slippage of surface brightening toward F3 is expected on its right (southern) side (indicated by a dashed arrow in Figure~\ref{fig:cartoon}c), but only weak signatures are seen in 304~{\AA} (bottom panels of Figure~\ref{fig:AIA304_slipping}). This is partly attributed to the reduced imaging contrast closer to the limb, with structures being `squeezed' into a narrower space by projection. On the other hand, the surface brightening is expected to be weak if open field lines result from the reconnections. These reconnections are termed ``slipping-like'', because it is impossible to determine whether the field lines slip through plasma at sub- or super-Alfv\'{e}nic speeds, which is termed slipping or slip-running reconnection \citep{Aulanier2006, Aulanier2007}.

\begin{figure}[ht!]
 \plotone{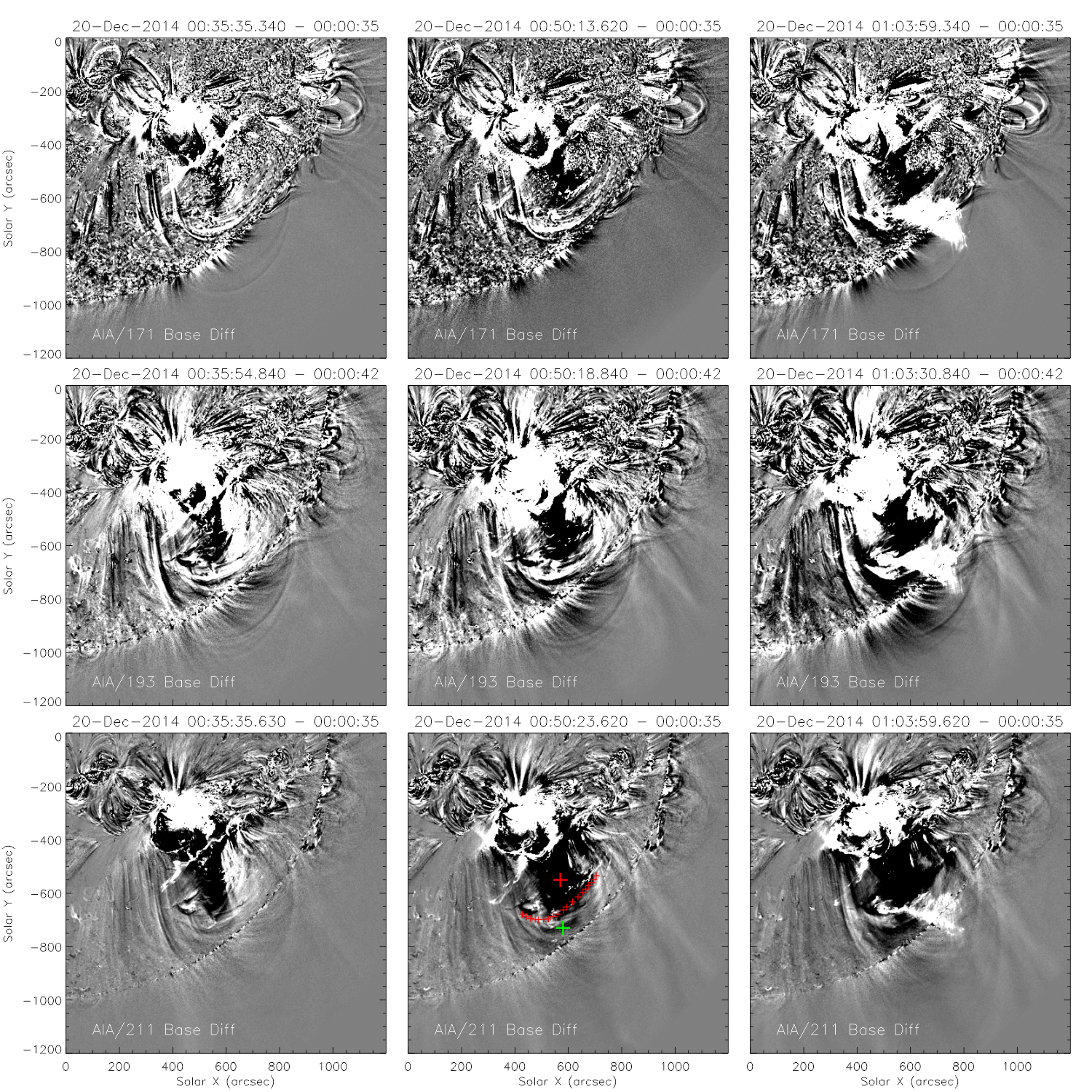}
 \caption{Coronal dimming in the sympathetic eruptions. The snapshots show base-difference images in 171~{\AA}, 193~{\AA}, and 211~{\AA}. An online animation is available showing the original, base- and running-difference images in the three AIA passbands. In the bottom-middle panel, the location of F3 is marked by `+' symbols, same as in Figure~\ref{fig:overview}a; the red (green) marks on the northern (southern) side of F3 the center of a $20''\times20''$ region, from which the light curves in Figure~\ref{fig:timeline}f are derived. \label{fig:AIA_base_diff}}
\end{figure}

As the slipping-like reconnections effectively open F3's overlying field, the coronal plasma above F3 would be evacuated, which predicts coronal dimming at both sides of F3 (Figure~\ref{fig:cartoon}c). This is indeed observed in AIA 171, 193, and 211~{\AA} before F3's eruption (Figures~\ref{fig:timeline}f and \ref{fig:AIA_base_diff}). Meanwhile, large-scale overlying loops would sense a temporary reduction in magnetic pressure above F3 due to the flux and mass evacuation, and hence contract toward F3 to seek a new equilibrium, as observed in AIA 171~{\AA} (Figures~\ref{fig:timeline}e and \ref{fig:loop}(d--f)). This so-called ``coronal implosion'' works in a wide range of flare phenomena \cite[e.g.,][]{Liu+Wang2009,Liu+Wang2010,Liu2009,Liu2012,Gou2017}. Most importantly, the slipping-like reconnections continually strengthen the magnetic force confining F2 while weakening the force confining F3 until F3 becomes unstable to the torus instability and erupts. 

The mechanism proposed here is not new, but to our knowledge signatures and effects of magnetic reconnection have seldom been identified in the literature on sympathetic eruptions, hence our results substantiate magnetic reconnection of large-scale field as a key causal link among sympathetic eruptions. These observations also highlight the ubiquity of magnetic reconnection by demonstrating its efficacy in the large-scale field of the quiet-Sun corona, where the magnetic field is weak and supposed to be approximately free of current, yet magnetic reconnection releases enough magnetic energy to heat up the chromosphere and newly reconnected coronal loops.

\acknowledgments D.W. acknowledges support by Natural Science Foundation of Anhui Province Education Department (KJ2017A493, KJ2017A491, gxyq2018030) and NSFC 11704003. R.L. acknowledges support by NSFC 41474151, 41774150, and 41761134088. Y.W. acknowledges support by NSFC 41774178 and 41574165. M.Z. acknowledges support by Natural Science Foundation of Anhui Province Education Department (KJ2016JD24) and Anhui Province Quality Engineering (2017zhkt161). This work is also supported by NSFC 41421063, CAS Key Research Program of Frontier Sciences QYZDB-SSW-DQC015, and the fundamental research funds for the central universities.

\software{SolarSoftWare \citep{Freeland+Handy2012}	}


\begin{thebibliography}{}
	\expandafter\ifx\csname natexlab\endcsname\relax\def\natexlab#1{#1}\fi
	
	\bibitem[{{Aulanier} {et~al.}(2006){Aulanier}, {Pariat}, {D{\'e}moulin}, \&
		{DeVore}}]{Aulanier2006}
	{Aulanier}, G., {Pariat}, E., {D{\'e}moulin}, P., \& {DeVore}, C.~R. 2006,
	\solphys, 238, 347
	
	\bibitem[{{Aulanier} {et~al.}(2007){Aulanier}, {Golub}, {DeLuca}, {Cirtain},
		{Kano}, {Lundquist}, {Narukage}, {Sakao}, \& {Weber}}]{Aulanier2007}
	{Aulanier}, G., {Golub}, L., {DeLuca}, E.~E., {et~al.} 2007, Science, 318, 1588
	
	\bibitem[{{Boerner} {et~al.}(2012){Boerner}, {Edwards}, {Lemen}, {Rausch},
		{Schrijver}, {Shine}, {Shing}, {Stern}, {Tarbell}, {Title}, {Wolfson},
		{Soufli}, {Spiller}, {Gullikson}, {McKenzie}, {Windt}, {Golub}, {Podgorski},
		{Testa}, \& {Weber}}]{Boerner2012}
	{Boerner}, P., {Edwards}, C., {Lemen}, J., {et~al.} 2012, \solphys, 275, 41
	
	\bibitem[{{D{\'e}moulin}(2006)}]{Demoulin2006}
	{D{\'e}moulin}, P. 2006, Advances in Space Research, 37, 1269
	
	\bibitem[{{D{\'e}moulin} \& {Aulanier}(2010)}]{Demoulin+Aulanier2010}
	{D{\'e}moulin}, P., \& {Aulanier}, G. 2010, \apj, 718, 1388
	
	\bibitem[{{Ding} {et~al.}(2006){Ding}, {Hu}, \& {Wang}}]{Ding2006}
	{Ding}, J.~Y., {Hu}, Y.~Q., \& {Wang}, J.~X. 2006, \solphys, 235, 223
	
	\bibitem[{{Freeland} \& {Handy}(2012)}]{Freeland+Handy2012}
	{Freeland}, S.~L., \& {Handy}, B.~N. 2012, {SolarSoft: Programming and data
		analysis environment for solar physics}, Astrophysics Source Code Library, ,
	, ascl:1208.013
	
	\bibitem[{{Gilbert} {et~al.}(2013){Gilbert}, {Inglis}, {Mays}, {Ofman},
		{Thompson}, \& {Young}}]{Gilbert2013}
	{Gilbert}, H.~R., {Inglis}, A.~R., {Mays}, M.~L., {et~al.} 2013, \apjl, 776,
	L12
	
	\bibitem[{Gou {et~al.}(2016)Gou, Liu, Wang, Liu, Zhuang, Chen, Zhang, \&
		Liu}]{Gou2016}
	Gou, T., Liu, R., Wang, Y., {et~al.} 2016, The Astrophysical Journal Letters,
	821, L28
	
	\bibitem[{{Gou} {et~al.}(2017){Gou}, {Veronig}, {Dickson}, {Hernandez-Perez},
		\& {Liu}}]{Gou2017}
	{Gou}, T., {Veronig}, A.~M., {Dickson}, E.~C., {Hernandez-Perez}, A., \& {Liu},
	R. 2017, \apjl, 845, L1
	
	\bibitem[{{Hudson} {et~al.}(1996){Hudson}, {Acton}, \& {Freeland}}]{Hudson1996}
	{Hudson}, H.~S., {Acton}, L.~W., \& {Freeland}, S.~L. 1996, \apj, 470, 629
	
	\bibitem[{{Jiang} {et~al.}(2008){Jiang}, {Shen}, {Yi}, {Yang}, \&
		{Wang}}]{Jiang2008}
	{Jiang}, Y., {Shen}, Y., {Yi}, B., {Yang}, J., \& {Wang}, J. 2008, \apj, 677,
	699
	
	\bibitem[{{Jiang} {et~al.}(2011){Jiang}, {Yang}, {Hong}, {Bi}, \&
		{Zheng}}]{Jiang2011}
	{Jiang}, Y., {Yang}, J., {Hong}, J., {Bi}, Y., \& {Zheng}, R. 2011, \apj, 738,
	179
	
	\bibitem[{{Jin} {et~al.}(2016){Jin}, {Schrijver}, {Cheung}, {DeRosa}, {Nitta},
		\& {Title}}]{Jin2016}
	{Jin}, M., {Schrijver}, C.~J., {Cheung}, M.~C.~M., {et~al.} 2016, \apj, 820, 16
	
	\bibitem[{{Joshi} {et~al.}(2016){Joshi}, {Schmieder}, {Magara}, {Guo}, \&
		{Aulanier}}]{Joshi2016}
	{Joshi}, N.~C., {Schmieder}, B., {Magara}, T., {Guo}, Y., \& {Aulanier}, G.
	2016, \apj, 820, 126
	
	\bibitem[{Kliem \& T{\"o}r{\"o}k(2006)}]{kliem2006torus}
	Kliem, B., \& T{\"o}r{\"o}k, T. 2006, Physical Review Letters, 96, 255002
	
	\bibitem[{Lemen {et~al.}(2011)Lemen, Akin, Boerner, Chou, Drake, Duncan,
		Edwards, Friedlaender, Heyman, Hurlburt, {et~al.}}]{lemen2011atmospheric}
	Lemen, J.~R., Akin, D.~J., Boerner, P.~F., {et~al.} 2011, in The Solar Dynamics
	Observatory (Springer), 17--40
	
	\bibitem[{{Liu} {et~al.}(2009{\natexlab{a}}){Liu}, {Lee}, {Karlick{\'y}},
		{Prasad Choudhary}, {Deng}, \& {Wang}}]{LiuC2009}
	{Liu}, C., {Lee}, J., {Karlick{\'y}}, M., {et~al.} 2009{\natexlab{a}}, \apj,
	703, 757
	
	\bibitem[{{Liu} {et~al.}(2012{\natexlab{a}}){Liu}, {Wang}, {Shen}, \&
		{Wang}}]{LiuK2012}
	{Liu}, K., {Wang}, Y., {Shen}, C., \& {Wang}, S. 2012{\natexlab{a}}, \apj, 744,
	168
	
	\bibitem[{{Liu} {et~al.}(2012{\natexlab{b}}){Liu}, {Kliem}, {T{\"o}r{\"o}k},
		{Liu}, {Titov}, {Lionello}, {Linker}, \& {Wang}}]{Liu2012dd}
	{Liu}, R., {Kliem}, B., {T{\"o}r{\"o}k}, T., {et~al.} 2012{\natexlab{b}}, \apj,
	756, 59
	
	\bibitem[{{Liu} {et~al.}(2012{\natexlab{c}}){Liu}, {Liu}, {T{\"o}r{\"o}k},
		{Wang}, \& {Wang}}]{Liu2012}
	{Liu}, R., {Liu}, C., {T{\"o}r{\"o}k}, T., {Wang}, Y., \& {Wang}, H.
	2012{\natexlab{c}}, \apj, 757, 150
	
	\bibitem[{{Liu} {et~al.}(2013){Liu}, {Liu}, {Xu}, {Liu}, {Kliem}, \&
		{Wang}}]{Liu2013}
	{Liu}, R., {Liu}, C., {Xu}, Y., {et~al.} 2013, \apj, 773, 166
	
	\bibitem[{{Liu} {et~al.}(2014){Liu}, {Titov}, {Gou}, {Wang}, {Liu}, \&
		{Wang}}]{Liu2014}
	{Liu}, R., {Titov}, V.~S., {Gou}, T., {et~al.} 2014, \apj, 790, 8
	
	\bibitem[{{Liu} \& {Wang}(2009)}]{Liu+Wang2009}
	{Liu}, R., \& {Wang}, H. 2009, \apjl, 703, L23
	
	\bibitem[{{Liu} \& {Wang}(2010)}]{Liu+Wang2010}
	---. 2010, \apjl, 714, L41
	
	\bibitem[{{Liu} {et~al.}(2009{\natexlab{b}}){Liu}, {Wang}, \&
		{Alexander}}]{Liu2009}
	{Liu}, R., {Wang}, H., \& {Alexander}, D. 2009{\natexlab{b}}, \apj, 696, 121
	
	\bibitem[{{Lynch} \& {Edmondson}(2013)}]{Lynch+Edmondson2013}
	{Lynch}, B.~J., \& {Edmondson}, J.~K. 2013, \apj, 764, 87
	
	\bibitem[{{Moon} {et~al.}(2003){Moon}, {Choe}, {Wang}, \& {Park}}]{Moon2003}
	{Moon}, Y.-J., {Choe}, G.~S., {Wang}, H., \& {Park}, Y.~D. 2003, \apj, 588,
	1176
	
	\bibitem[{{Pesnell} {et~al.}(2012){Pesnell}, {Thompson}, \&
		{Chamberlin}}]{Pesnell2012}
	{Pesnell}, W.~D., {Thompson}, B.~J., \& {Chamberlin}, P.~C. 2012, \solphys,
	275, 3
	
	\bibitem[{{Schatten} {et~al.}(1969){Schatten}, {Wilcox}, \&
		{Ness}}]{Schatten1969}
	{Schatten}, K.~H., {Wilcox}, J.~M., \& {Ness}, N.~F. 1969, \solphys, 6, 442
	
	\bibitem[{Scherrer {et~al.}(2012)Scherrer, Schou, Bush, Kosovichev, Bogart,
		Hoeksema, Liu, Duvall, Zhao, Schrijver, {et~al.}}]{scherrer2012helioseismic}
	Scherrer, P.~H., Schou, J., Bush, R., {et~al.} 2012, Solar Physics, 275, 207
	
	\bibitem[{Schou {et~al.}(2011)Schou, Scherrer, Bush, Wachter, Couvidat,
		Rabello-Soares, Bogart, Hoeksema, Liu, Duvall, {et~al.}}]{schou2011design}
	Schou, J., Scherrer, P., Bush, R., {et~al.} 2011, in The Solar Dynamics
	Observatory (Springer), 229--259
	
	\bibitem[{Schrijver \& DeRosa(2003)}]{schrijver2003photospheric}
	Schrijver, C.~J., \& DeRosa, M.~L. 2003, Solar Physics, 212, 165
	
	\bibitem[{{Schrijver} \& {Title}(2011)}]{Schrijver+Title2011}
	{Schrijver}, C.~J., \& {Title}, A.~M. 2011, Journal of Geophysical Research
	(Space Physics), 116, A04108
	
	\bibitem[{{Schrijver} {et~al.}(2013){Schrijver}, {Title}, {Yeates}, \&
		{DeRosa}}]{Schrijver2013}
	{Schrijver}, C.~J., {Title}, A.~M., {Yeates}, A.~R., \& {DeRosa}, M.~L. 2013,
	\apj, 773, 93
	
	\bibitem[{{Shen} {et~al.}(2012){Shen}, {Liu}, \& {Su}}]{Shen2012}
	{Shen}, Y., {Liu}, Y., \& {Su}, J. 2012, \apj, 750, 12
	
	\bibitem[{{Sun} {et~al.}(2015){Sun}, {Hoeksema}, {Liu}, \& {Zhao}}]{Sun2015}
	{Sun}, X., {Hoeksema}, J.~T., {Liu}, Y., \& {Zhao}, J. 2015, \apj, 798, 114
	
	\bibitem[{{Titov} {et~al.}(2012){Titov}, {Mikic}, {T{\"o}r{\"o}k}, {Linker}, \&
		{Panasenco}}]{Titov2012}
	{Titov}, V.~S., {Mikic}, Z., {T{\"o}r{\"o}k}, T., {Linker}, J.~A., \&
	{Panasenco}, O. 2012, \apj, 759, 70
	
	\bibitem[{T{\"o}r{\"o}k \& Kliem(2007)}]{torok2007numerical}
	T{\"o}r{\"o}k, T., \& Kliem, B. 2007, Astronomische Nachrichten, 328, 743
	
	\bibitem[{{T{\"o}r{\"o}k} {et~al.}(2011){T{\"o}r{\"o}k}, {Panasenco}, {Titov},
		{Miki{\'c}}, {Reeves}, {Velli}, {Linker}, \& {De Toma}}]{Torok2011}
	{T{\"o}r{\"o}k}, T., {Panasenco}, O., {Titov}, V.~S., {et~al.} 2011, \apjl,
	739, L63
	
	\bibitem[{Wang {et~al.}(2017)Wang, Liu, Wang, Liu, Chen, Liu, Zhou, \&
		Zhang}]{wang2017critical}
	Wang, D., Liu, R., Wang, Y., {et~al.} 2017, The Astrophysical Journal Letters,
	843, L9
	
	\bibitem[{{Wang} {et~al.}(2001){Wang}, {Chae}, {Yurchyshyn}, {Yang},
		{Steinegger}, \& {Goode}}]{Wang2001}
	{Wang}, H., {Chae}, J., {Yurchyshyn}, V., {et~al.} 2001, \apj, 559, 1171
	
	\bibitem[{{Wang} {et~al.}(2016){Wang}, {Liu}, {Zimovets}, {Hu}, {Dai}, \&
		{Yang}}]{Wang2016}
	{Wang}, R., {Liu}, Y.~D., {Zimovets}, I., {et~al.} 2016, \apjl, 827, L12
	
	\bibitem[{{Wang} {et~al.}(2007){Wang}, {Sheeley}, \& {Rich}}]{Wang2007}
	{Wang}, Y.-M., {Sheeley}, Jr., N.~R., \& {Rich}, N.~B. 2007, \apj, 658, 1340
	
	\bibitem[{{Xu} {et~al.}(2010){Xu}, {Jing}, \& {Wang}}]{Xu2010}
	{Xu}, Y., {Jing}, J., \& {Wang}, H. 2010, \solphys, 264, 81
	
	\bibitem[{{Yang} {et~al.}(2012){Yang}, {Jiang}, {Zheng}, {Bi}, {Hong}, \&
		{Yang}}]{Yang2012}
	{Yang}, J., {Jiang}, Y., {Zheng}, R., {et~al.} 2012, \apj, 745, 9
	
	\bibitem[{{Zuccarello} {et~al.}(2009){Zuccarello}, {Romano}, {Farnik},
		{Karlicky}, {Contarino}, {Battiato}, {Guglielmino}, {Comparato}, \&
		{Ugarte-Urra}}]{Zuccarello2009}
	{Zuccarello}, F., {Romano}, P., {Farnik}, F., {et~al.} 2009, \aap, 493, 629
	
\end{thebibliography}



\end{document}